# Ultranarrow Bright Single-Photon Emitters in Diamond with Strong Broadband Phonon Decoupling

Swetapadma Sahoo[1-4], Péter Udvarhelyi[5], Jaden Li[1], Darwon Kim[1], Viatcheslav Agafonov[6], Valery A. Davydov[7], Benjamin Lawrie[8,9], Prineha Narang[10,11], Simeon I. Bogdanov[1-4,*]

**Affiliations:**

[1]Department of Electrical and Computer Engineering, University of Illinois at Urbana-Champaign, Urbana, Illinois 61801, USA

[2]Nick Holonyak, Jr. Micro and Nanotechnology Laboratory, University of Illinois at Urbana-Champaign, Urbana, Illinois 61801, USA

[3]Illinois Quantum Information Science and Technology Center, University of Illinois Urbana-Champaign, Urbana, Illinois 61801, USA

[4]The Grainger College of Engineering, University of Illinois Urbana-Champaign, Urbana, IL, USA

[5]Department of Chemistry and Biochemistry, University of California Los Angeles, Los Angeles, California 90095, United States

[6]GREMAN, CNRS, UMR 7347, INSA CVL, Université de Tours, 37200 Tours, France

[7]L.F. Vereshchagin Institute for High Pressure Physics, Russian Academy of Sciences, Troitsk, Moscow, 108840, Russia

[8]Materials Science and Technology Division, Oak Ridge National Laboratory, Oak Ridge, TN 37831, USA

[9]Center for Nanophase Materials Sciences, Oak Ridge National Laboratory, Oak Ridge, TN 37831, USA

[10]Division of Physical Sciences, College of Letters and Science, University of California Los Angeles, Los Angeles, California 90095, United States

[11]Department of Electrical and Computer Engineering, University of California Los Angeles, Los Angeles, California 90095, United States

*Corresponding author. Email: bogdanov@illinois.edu

**Abstract:** Single-photon emitters are fundamental building blocks for quantum information processing, communication and sensing. However, unwanted interactions with bulk phonons in their host environment strongly limit their coherence and controllability. We report single color centers in nanodiamonds that are strongly and comprehensively decoupled from the bulk phononic environment. The color centers feature record-narrow linewidths down to 0.3 nm at room temperature and stable, bright emission, exceeding 10 Mcps in saturation. Notably, the bulk phonon sideband is almost entirely suppressed, revealing the presence of a single localized vibrational mode outside the diamond phonon band. Our observations and simulations point towards a unique mechanism for phonon decoupling in common wide-gap materials, based on a strongly radiative orbital transition coupled to a localized vibrational mode. The new color center enables qualitatively higher performance for applications in quantum networks and nanoscale sensing, and the exploration of new physical resources associated with vibrational states.

## Introduction:

Color centers in wide-gap materials[1] are solid-state equivalents of single atoms, featuring an atom-like structure of quantum levels, with coupled electronic orbital, electronic spin, and nuclear spin transitions. Harnessing such quantized transitions in color centers enables a range of quantum functionalities, including quantum registers[2], memories[3,4], sensors[5], and single-photon sources[1]. Owing to their solid-state nature, color centers can be interfaced with on-chip optical, electrical, and RF control channels, as well as external electronic and photonic devices, forming highly functional quantum circuits[6–9]. While single atoms are naturally identical and highly coherent, solid-state host materials affect the properties of color centers significantly, hampering their scalable deployment. In particular, the unwanted interactions of electronic orbital states with phonons create numerous challenges (Fig. 1(a)). First, phonon dephasing of the zero-phonon line (ZPL) leads to photon distinguishability and affects the fidelity of optical spin control operations. Second, phonon emission during the radiative transition routes photons into the undesirable bulk phonon side band (PSB). Third, phonon-mediated non-radiative transitions reduce the overall quantum yield. These phonon interactions generally require the color centers to operate at deep cryogenic temperatures, usually in tight integration with individually tailored nanophotonic resonators that enhance the coupling to specific optical modes.

For decades, the nitrogen-vacancy (NV) center in diamond, offering a coherent and optically active spin, has been the benchmark for all color center technologies. Unfortunately, the NV center lacks inversion symmetry due to a substitutional nitrogen impurity, and its emission is impractically broadband for most quantum information applications. A major advance was achieved with the introduction of group IV split-vacancy centers (G4V)[10]: SiVs[11,12], GeVs[13,14], SnVs[15,16] and PbVs[17,18]. In these centers, an interstitial impurity atom is located between two half-vacancy sites, and the phonon coupling is significantly reduced due to inversion symmetry. In recent years, several groups have reported color centers, including silicon-related centers in nanodiamonds on Iridium substrate[19], L1[20], and LX centers[21], featuring unusually narrow linewidths and strongly suppressed phonon sidebands. These observations hinted at the existence of a new class of color centers that are more strongly decoupled from phonons than the G4V centers. However, the origin of this phonon decoupling and its potential have remained unclear.

Here, we report a new color center, hereafter referred to as the IL1 center, in low-strain nanodiamonds (NDs), exhibiting exceptional and comprehensive decoupling from bulk phonons. This decoupling is characterized by record-narrow room-temperature ZPL linewidth, a multi-order suppression of the bulk PSB, and a high quantum yield, leading to exceptionally bright single-photon emission (Fig. 1(a)). These properties are consistently observed across a large set of emitters. By studying the temperature dependence of the ZPL linewidth, we show that IL1 color centers hold the potential for producing indistinguishable photons at non-cryogenic temperatures. Remarkably, the orbital transition is coupled to a single localized vibrational mode (LVM), contributing to the isolation from bulk phonons. The coupling of a single orbital transition to a single LVM oscillator unobscured by other dynamics creates a "laboratory" setting for studying coherent properties of single vibrons as a potential resource for hybrid quantum systems. Finally, we propose an atomic origin of the IL1 color center supported by first-principles calculations and discuss the physical mechanism of decoupling from bulk phonons, opening avenues to designing phonon-isolated defects in diamond and other wide-gap materials.

# Results:

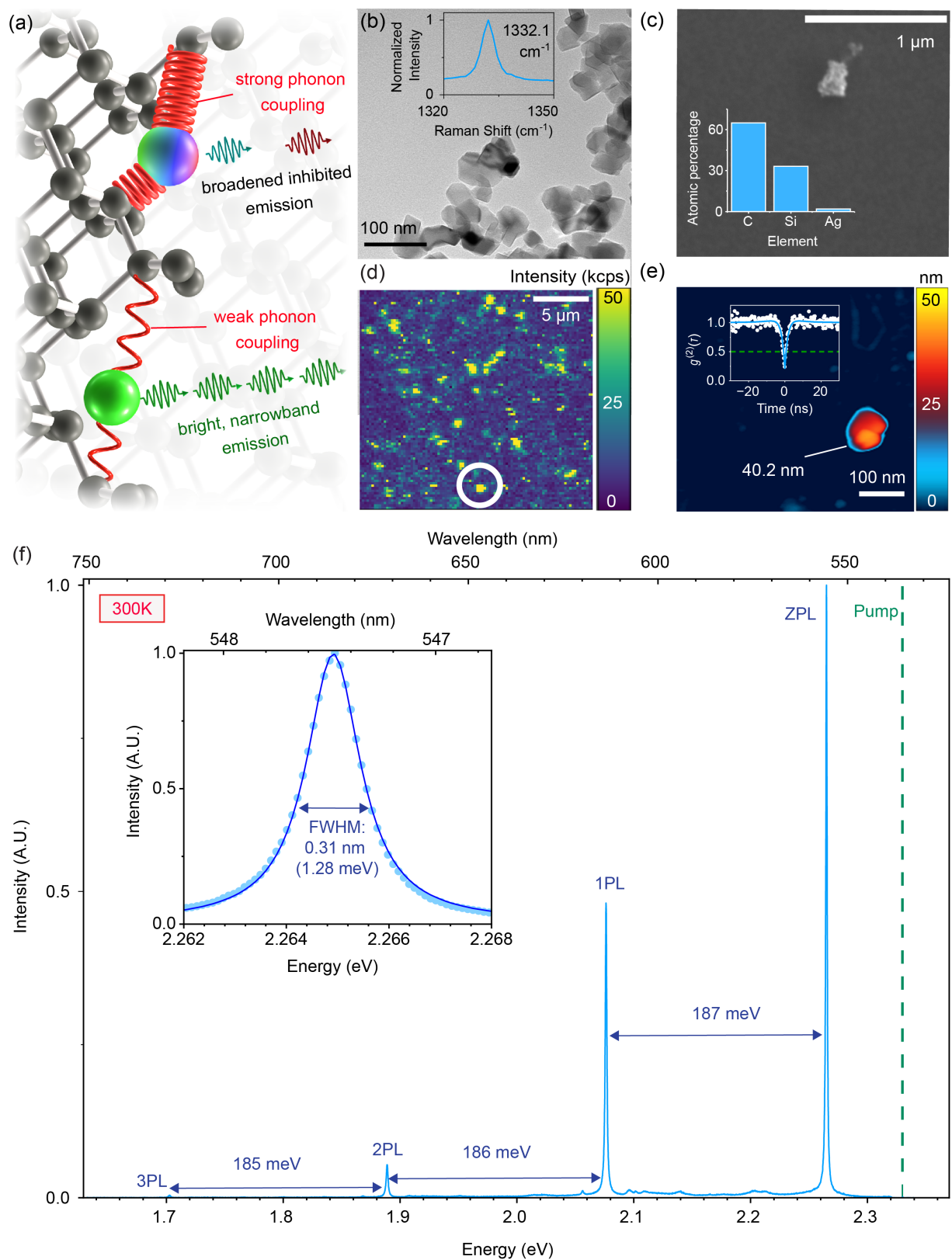

**Fig. 1.** Characterization of IL1 centers in nanodiamonds (NDs) with strong phonon decoupling at room temperature. (a) Coupling to bulk phonons broadens the ZPL emission and creates undesired decay pathways in color centers

(above). Reducing the phonon coupling enables narrowband and bright single-photon emission (below). (b) A TEM image of HPHT-synthesized diamond powder reveals clusters of sub-100 nm NDs. Inset shows the Raman spectrum with a characteristic diamond peak at 1332.1 cm$^{-1}$. (c) EDS analysis of NDs dispersed on an Ag-coated Si substrate confirms that the particles are predominantly composed of carbon. (d) A PL intensity map of a 20 μm × 20 μm area containing dispersed NDs features isolated IL1 centers. The white circle marks the location of a representative emitter characterized in (e). (e) AFM image of the ND host of the emitter circled in (d). The inset shows the corresponding second-order autocorrelation measurement. (f) The background-corrected room-temperature PL spectrum of a representative IL1 center reveals four characteristic narrow peaks.

The NDs studied in this work are synthesized by high-pressure high-temperature (HPHT) treatment of a homogeneous mixture of 1-fluoroadamantane, $C_{10}H_{15}F$ (FAD) and triphenylchlorosilane, $C_{18}H_{15}ClSi$ (see Methods). First, we verify that the sample indeed consists of nanoscale diamond particles. A typical transmission electron microscopy (TEM) image of the synthesized nanoparticles is shown in Fig. 1(b). Individual nanoparticles exhibit linear sizes of 43 ± 7 nm. The Raman spectrum of the synthesized powder features a distinct Raman peak at 1332.1 cm$^{-1}$ characteristic of the diamond sp$^3$ lattice (Fig. 1(b) inset). Energy-dispersive X-ray spectroscopy of nanoparticles dropcasted on a thin silver film deposited on silicon confirms their carbon-dominated elemental composition (Fig. 1(c)). Photoluminescence (PL) intensity maps (Fig. 1(d)) reveal isolated emitters, many of which are hosted by NDs with sizes of a few tens of nanometers (Fig. 1(e)) and exhibit single-photon emission confirmed through second-order correlation measurements (Fig. 1(e) inset).

Notably, most single-photon emitters observed in this sample feature a characteristic spectrum with four sharp, nearly harmonically spaced emission lines. A representative spectrum is shown in Fig. 1(f). These characteristic four-peak spectra are systematically accompanied by PL antibunching ($g^2(0) < 0.5$) and can therefore be regarded as spectral fingerprints of single IL1 centers. The emission peak with the highest energy is also the narrowest, consistent with the properties expected from a zero-phonon line (ZPL). At room temperature, the ZPL peak's linewidth of 0.31 nm (1.28 meV) at 547.5 nm (2.27 eV), is significantly narrower than that observed in known color centers. All four peaks have a practically symmetric Lorentzian lineshape, indicating that the linewidth is dominated by homogenous broadening with no noticeable spectral diffusion. Remarkably, the bulk PSB, typically dominating the emission of color centers at room temperature, is essentially absent, indicating a strong suppression of the electronic transition's coupling to diamond bulk phonon modes. The peaks are nearly harmonically distributed with an energy difference of about 187 meV, which is above the optical phonon energy in diamond (165 meV). This large energy separation indicates that a single localized vibrational mode is responsible for the four-peak vibronic spectrum [19,22].

Next, we study the photophysical properties of a single IL1 center. The emission of IL1 centers is exceptionally bright, with saturation intensities reaching up to 12.5 Mcps after background correction (Fig. 2 (a)). Saturation curves for additional single emitters are shown in Supplementary Information S1. The curves are fitted with the equation $I = I_\infty P/(P_{sat} + P)$, where $I_\infty$ is the maximum emission intensity and $P_{sat}$ is the saturated power. Notably, there are no signs of blinking down to the millisecond time scale (Fig. 2 (a) inset) for this emitter, consistently with the data obtained on other single IL1 centers (see Supplementary Information S2). Time-resolved measurements of the same emitter (Fig. 2(b)) reveal no significant "bunching" behavior at $|\tau| > 0$. We measure an excited-state lifetime of 2.4 ± 0.1 ns (Fig. 2(c), which is consistent with other IL1 emitters (see Supplementary Information S3). Assuming a setup efficiency of 3.8%, we find the CW quantum yield to be 92% in the two-level approximation (Supplementary Information S4), justified by the absence of significant "bunching" signatures in the autocorrelation curves. This high quantum yield means that the excited state decay has a strong radiative component. Given that the nanodiamond hosts extend the radiative lifetime by a factor of up to 10 compared to bulk diamond lifetimes[23,24], the radiative rate component is among the fastest found in color centers and on par with those found in quantum dots.

The polarization diagram $I(\theta)$ of the emitter is shown in Fig. 2(d). The degree of linear polarization is given by polarization visibility, $V = (I_{max} - I_{min})/(I_{max} + I_{min}) = 67\%$. Other emitters feature a visibility of up to 91% (see Supplementary Information S5), indicating that the emission originates from a single linearly polarized dipole transition [25]. Some loss in visibility can be attributed to the random orientation of NDs, depolarization effects in optical components such as the dichroic mirror, and loss of polarization contrast due to imaging with a NA = 1.49 oil immersion objective[26].

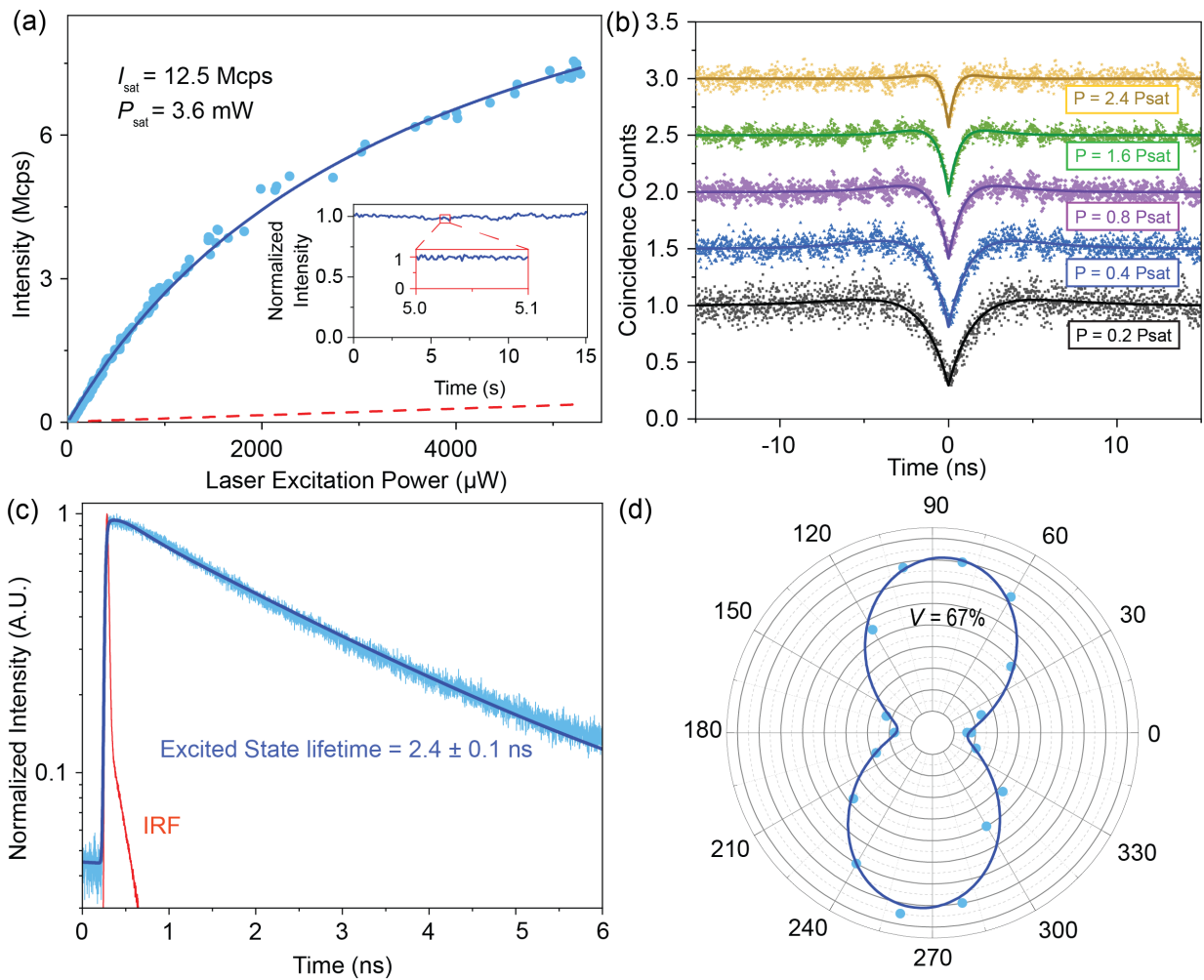

**Fig. 2**. Photophysical properties of a single IL1 center (a) Saturation curve (blue), corrected by subtracting the background component (dashed red), yields a saturation power of $P_{sat}$ = 3.6 mW and saturation intensity $I_\infty$ = 12.5 × $10^6$ photon counts/s (cps). The inset shows stable photon emission with no blinking. (b) Power-dependent autocorrelation histograms for the emitter, given in terms of its saturation power $P_{sat}$ in (a), exhibit negligible bunching. (c) Excited state lifetime measurement, fitted with a double exponential (blue line) against the instrument response function (IRF, red). (d) Emission polarization diagram $I(\theta)$, fitted with a $\sin^2(\theta)$-form fit function, shows a polarization visibility of 67%.

To gain better insight into the electron–phonon interactions, we study the temperature-dependent PL spectra of IL1 centers. The spectral features remain consistent at all temperatures, with all four peaks persisting until 4K. At 125 K, the ZPL linewidth reaches the spectrometer's resolution limit of 0.047 nm, as shown in Fig. 3(a) (Princeton Instruments IsoPlane SCT-320 spectrograph, coupled with a Pixis 400BR Excelon camera and a 2400 grooves mm$^{-1}$ grating). Furthermore, no peak splitting is observed down to the resolution limit of the instrument, and no indication of a bulk PSB is detected above the noise floor up to 300 K (see Supplementary Information S6). We observe a $T^3$ scaling of the ZPL FWHM above 200K (Fig. 3(b)), consistent with previously reported trends for several other diamond color centers[27,28]. The temperature dependence of 1PL and 2PL peak FWHMs is consistent with $\alpha T^3 + \gamma_{vib}$ and $\alpha T^3 + 2\gamma_{vib}$, where $\gamma_{vib}$= 0.26 meV is attributed to the localized vibrational state's decay rate. The temperature-dependent optical lineshift is consistent with an $aT^4 + bT^2$ scaling (Fig.3(c)). Such lineshifts have been observed in previous reports for SiV[27], NV, and telecom wavelength centers[29] in diamond, where they were attributed to the combination of electron-phonon coupling, diamond lattice expansion, and softening of elastic bonds in the excited states.

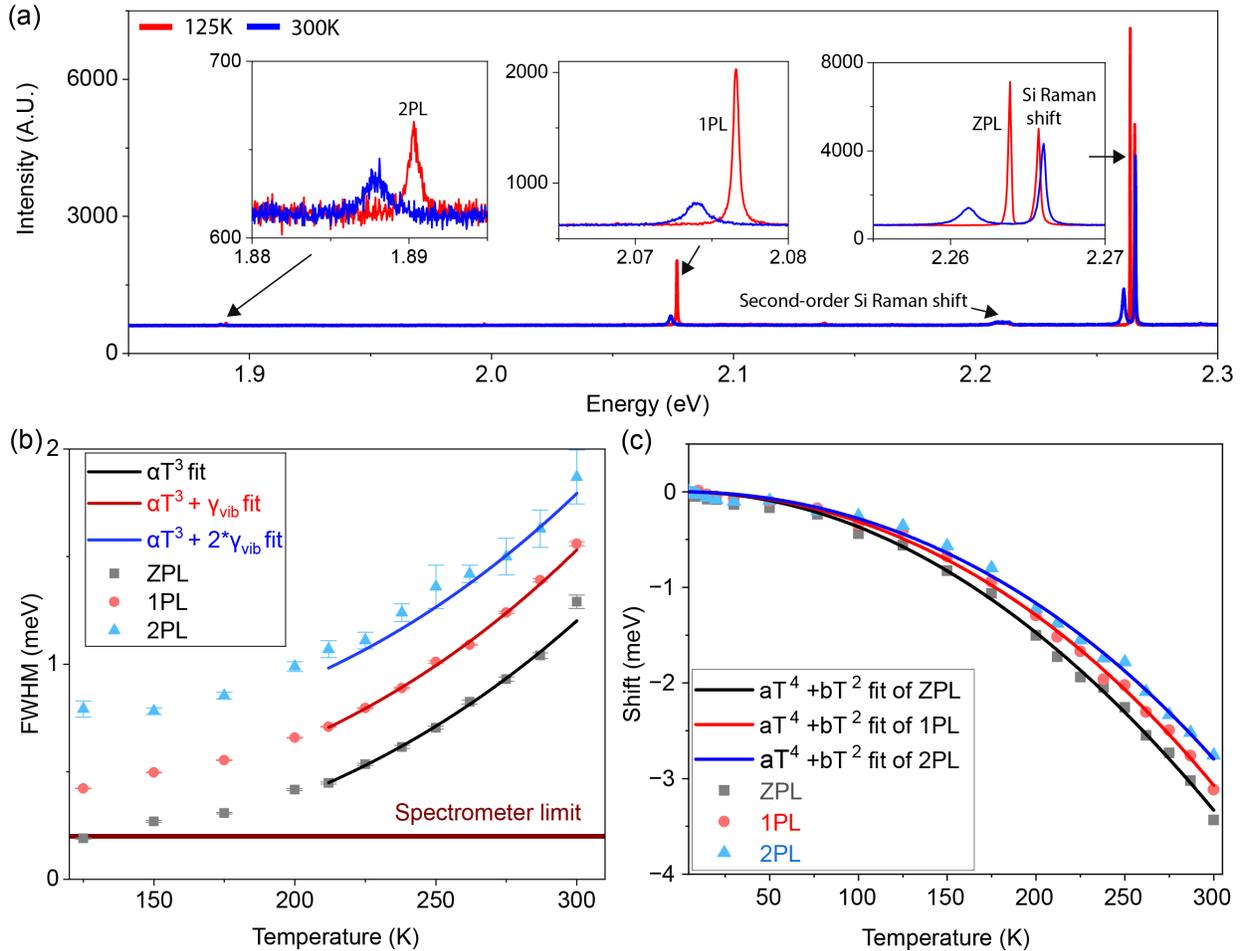

**Fig. 3.** Analysis of the temperature dependence of the PL spectra of IL1 centers in NDs dispersed on a Si substrate. (a) Emission spectra at 125 K and 300 K, without correction. (b) Temperature dependence of the FWHM of ZPL, peak 1PL and 2PL from 125 K to 300 K, fitted with $\alpha T^3$, $\alpha T^3 + \gamma_{vib}$ and $\alpha T^3 + 2\gamma_{vib}$, respectively. (c) Temperature dependence of the wavelength shift of ZPL, peaks 1PL and 2PL from 4.4K to 300K, fitted with $aT^4 + bT^2$.

Statistical analysis reveals low residual strain in the NDs, uniform LVM and ZPL energies, and consistently narrow ZPL linewidths at room temperature. NDs exhibited low residual strain, a characteristic of adamantane-assisted synthesis[13,30]. We observe an average Raman line peak energy of 1331.96 ± 0.21 cm$^{-1}$ (Fig. 4(a)), redshifted from the value of 1332.5 cm$^{-1}$ ($v_0$) for unstrained diamond[31], and indicating a slight residual tensile strain, an order of magnitude weaker than that reported in CVD-grown nanodiamonds[32]. This residual strain can be used as a lower-bound estimate for the strain in the individual NDs. The Raman line FWHMs range from 3.7 to 5.4 cm$^{-1}$. Part of this broadening may originate from phonon confinement effects in the nanodiamonds[33]. Nonetheless, it provides an upper bound for the characteristic nanoscale strain within individual particles. These two bounds give us a strain range of 0.02 – 0.18% calculated via $\text{stress} = 0.34\,\text{GPa}/\text{cm}^{-1}\left(v - 1332.5\,\text{cm}^{-1}\right)$ where $v$ is the measured Raman frequency and the Young's modulus of diamond of 1100 GPa.

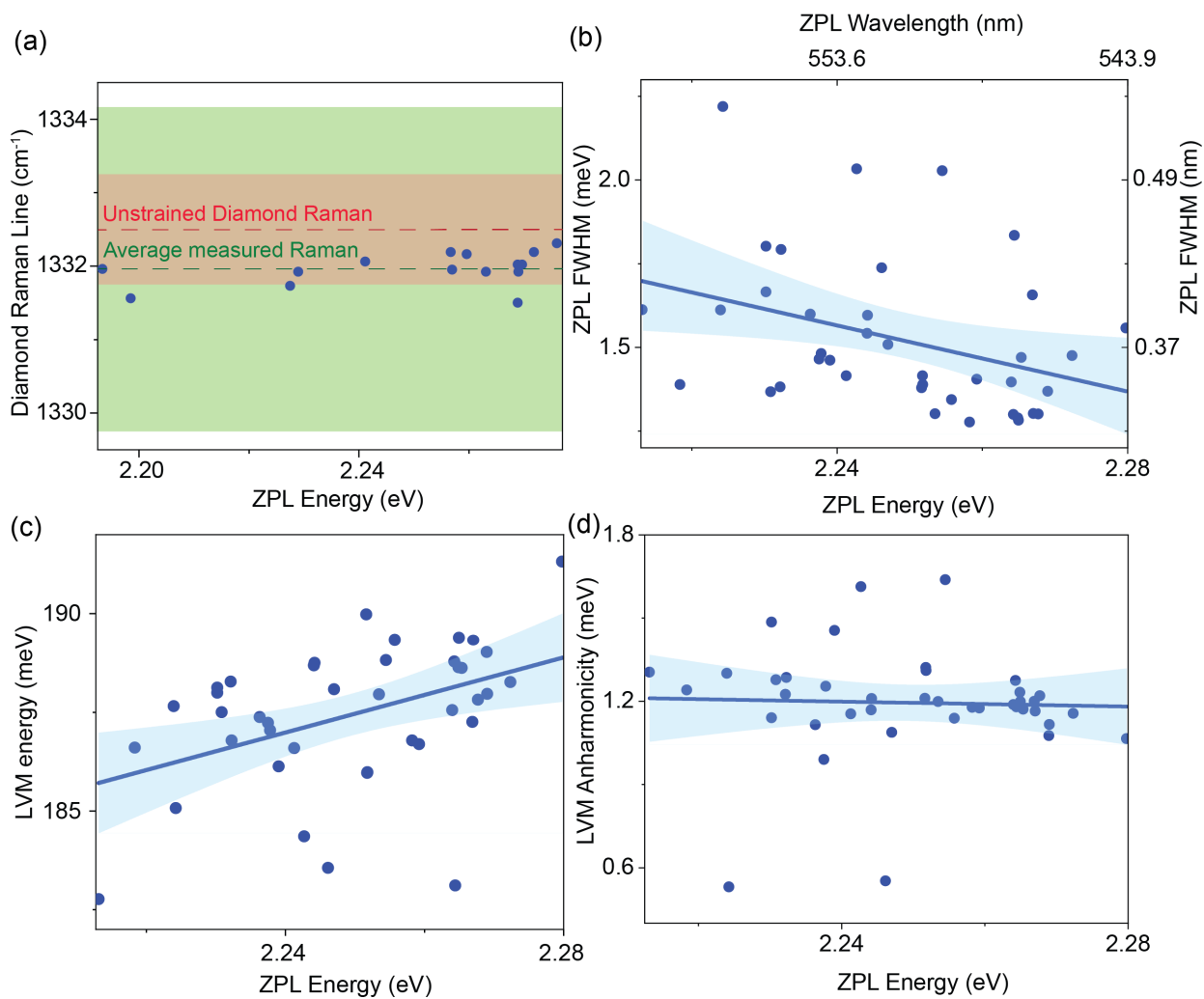

**Fig. 4.** Statistical analysis of IL1 photophysical properties at room temperature. (a) The diamond Raman lines in IL1-NDs consistently show a downshift from the unstrained diamond Raman line of 1332.5 cm$^{-1}$ (red dashed line), with an average Raman line observed at 1331.96 cm$^{-1}$ (green dashed line). The red shaded band shows the FWHM of an unstrained diamond Raman line (1.5 cm$^{-1}$) and the green band shows the average FWHM of IL1 Raman lines (4.4 cm$^{-1}$). (b) ZPL FWHM as a function of ZPL energy. (c) LVM energy as a function of ZPL energy. (d) LVM anharmonicity as a function of ZPL energy. Linear regression to all datapoints is shown with solid lines. The blue shaded areas indicate the 95% confidence intervals of the regression.

Among 40 probed emitters, each exhibiting four characteristic peaks, the ZPL wavelengths (Fig. 4(b)) lie in the range between 544 and 560 nm, with an average ZPL energy of 2.25 eV and a standard deviation of 0.016 eV. The peaks 1PL, 2PL, and 3PL are similarly distributed (see Supplementary Information S7). Remarkably, all the color centers have a ZPL linewidth below 0.6 nm (with average ZPL linewidth ~ 0.37 nm (1.5 meV) (Fig. 4(b)). We additionally observe a trend of ZPL linewidths becoming broader at longer wavelengths. A similar trend has been reported for the narrowband Si-related centers in nanodiamonds[32], resulting from the strain-induced modification of electron-phonon coupling.

We further observe a trend of increasing LVM energy with ZPL energy, shown Fig. 4(c). The average linewidths of 1PL, 2PL, and 3PL peaks are 0.5 nm (1.8 meV), 0.7 nm (2.1 meV), and 1 nm (2.4 meV) (see Supplementary Information S6). The linewidth increases with each peak by approximately the vibrational decay rate $\gamma_{vib}$. Furthermore, the vibrational spectral peaks consistently feature a slight anharmonicity of ~1.2 ± 0.2 meV/level (Fig. 4(d)). This anharmonicity contributes to a slight asymmetry of ZPL[34], as observed in the insert of Fig. 1(g).

**Atomic structure simulations:**

To provide candidate structures for the IL1 center, we systematically evaluate the defect configurations feasible during ND synthesis, searching for matching properties. Regarding the element composition, we first consider the incorporation of H and Si atoms, while excluding the volatile halogen elements present in the growth system. As optical activity requires the presence of several localized levels inside the bandgap, vacancy-related and interstitial complex defects are ideal candidates. The most important fingerprint for identification is the high-energy LVM of the defect. The observation of a high-energy LVM is in contrast to the phonon softening effect associated with the expansive strain environment created by the vacancy defect. Indeed, our calculations prove that G4V defects cannot possess large-energy LVMs (see Fig. 5(a)). Instead, two other origins are possible: the bending vibrational mode of the H-C bond, or vibrations of a reconstructed C-C bond in a compressive local environment, such as that associated with interstitial defects. As the role of hydrogen in vacancy[35], and in silicon-vacancy defects[36] has been studied in great detail without showing matching optical signatures to the IL1 center, we focus on self-interstitial defects and their aggregation. At least three such defects feature LVMs in the correct energy range (see Fig. 5(b), (c), (d), (e)). The first candidate is a single carbon interstitial defect forming a [001] split-interstitial [37]. Because of the high $D_{2d}$ symmetry, the two dangling bonds form a double-degenerate level inside the bandgap, which is optically inactive. The two-interstitial aggregate $C_iC_i$ of $C_{2h}$ symmetry is also optically inactive for the same reasons. Another aggregate with low formation energy is the $C_iCC_i$ second neighbor interstitial complex in $C_{2v}$ symmetry. The lowered symmetry splits the orbital degeneracy. However, this splitting is not strong enough to result in a large-energy ZPL transition. Further increasing the number of interstitial carbon atoms to 3, the lowest energy structures have the flexibility to mend all the dangling bonds, resulting in optically inactive defects. Instead, we investigate the metastable $C_iC_iC_i$ compact defect aggregated along the [110] direction in the lattice. This configuration is promising as structural relaxations are expected along the same direction, which is compatible with coupling to the high-energy, rocking LVM mode of the interstitial defect. Our calculations show that the formation energy of the single split interstitial defect (11.3 eV) is larger than any of the above-mentioned aggregations normalized by the number of interstitial atoms. The formation energy for the candidate $C_iC_iC_i$ defect is 9.9 eV. These results show that self-interstitials preferentially aggregate in diamond when no competing annihilation mechanisms are available.

Once formed, the dissociation of the metastable $C_iC_iC_i$ defect is hindered by an energy barrier of 1.3 eV according to our reaction path calculations (see Methods).

The calculations on the $C_iC_iC_i$ defect in the neutral charge state result in a singlet spin state, with a metastable triplet 1.1 eV above. The occupied and unoccupied orbitals are located relatively close to the respective band edges, giving a large region of stability for the neutral charge state (see Fig. 6(a)). The localization of the orbital wavefunction is very similar for the highest occupied and lowest unoccupied levels. This suggests that the optical transition dipole is large, while the macroscopic dipole change is small, leading to a strong and stable optical emission. This lowest energy transition produces ZPL at 2.083 eV. The underestimation compared to the experimental IL1 line can be mainly attributed to the many-body correlation in the singlet states, which is not accounted for in our density functional theory calculations. We estimate the optical lifetime of 5.3 ns from the calculated transition dipole moment, which matches the longest lifetimes recorded in the experiment (see Supplementary Information S3). Consistent with the PL polarization data, the $C_iC_iC_i$ ZPL belongs to a single dipole transition between clearly separated orbital singlets with an electric dipole along the [110] direction.

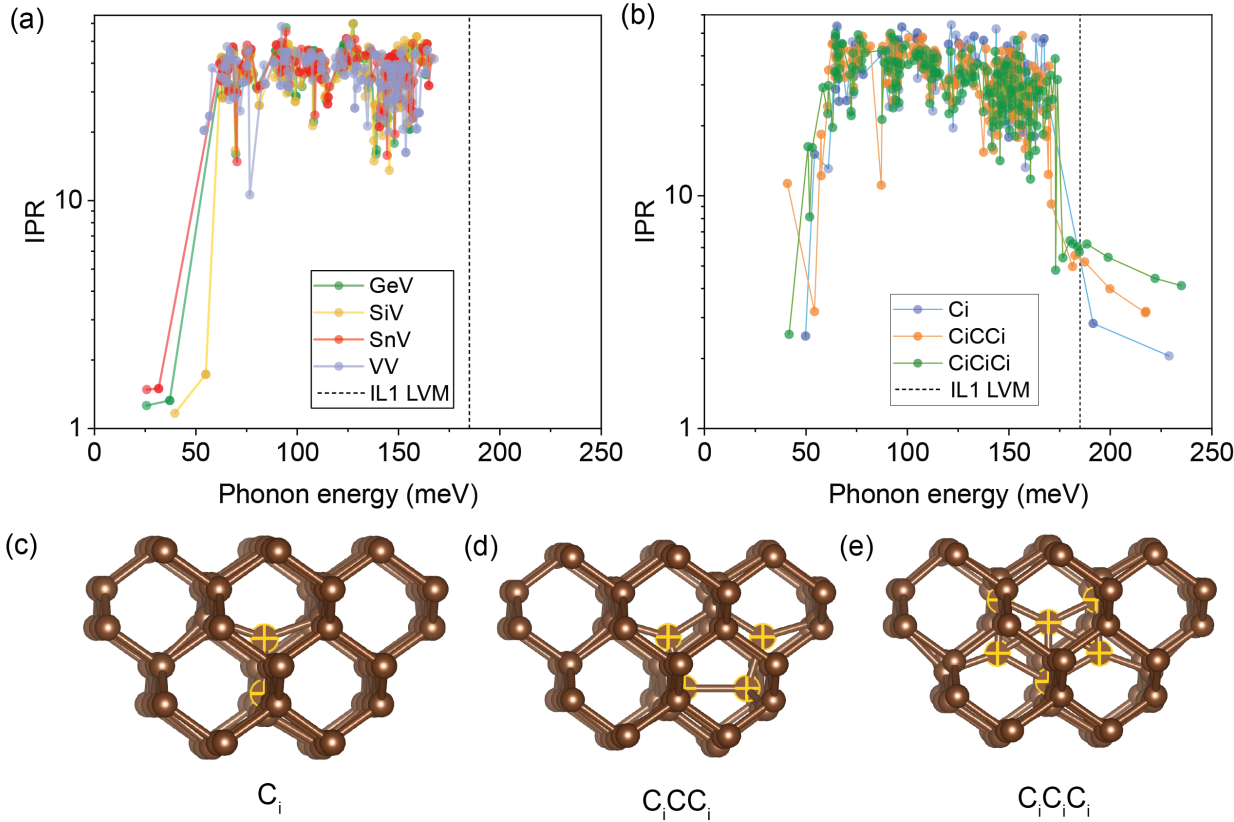

**Fig. 5.** Inverse participation ratio (IPR) of the phonon modes calculated for various interstitial defect structures. Localized vibration modes (LVMs) are associated with the lowest IPR values. While vacancy-related structures generally do not support LVMs of large energy (a), we find several self-interstitial complexes possessing LVMs around the experimental IL1 LVM line (b). Such defect structures in the diamond lattice are visualized below the plots (c,d,e).

The analysis of the ZPL-strain coupling shows a susceptibility of nearly 6 eV/strain, a relatively strong susceptibility which is due to the large intrinsic strain of the interstitial complex. This large susceptibility can partially explain the observed inhomogeneous ZPL energy distribution.

We also simulate the emission sideband based on the ionic relaxation upon excitation. In Fig. 6 (b), the 186 meV LVM mode shows a very strong coupling to the electronic transition with the largest partial Huang-Rhys (S) factor in the system. The total S factor is 2.4 and the corresponding Debye-Waller factor is 9%, lower than that observed in the experiment (38%). While the point symmetry and single orbital state of proposed structure does not support classic Jahn–Teller effects [38,39], pseudo–Jahn–Teller vibronic interactions [40] may still be present in the IL1 center. These interactions generally weaken with large energy separation, as is the case here, so the corrections are expected to be small. Nonetheless, such interactions can give corrections beyond the Huang–Rhys framework used to simulate our PL spectrum. In addition, confinement and surface effects can further modify the spectrum relative to the bulk model simulations[41]. However, most of the calculated optical properties of the $C_iC_iC_i$ defect are in excellent agreement with our experiments. Therefore, the $C_iC_iC_i$ (or a similar self-interstitial aggregate) constitutes a viable candidate for the origin of the IL1 center.

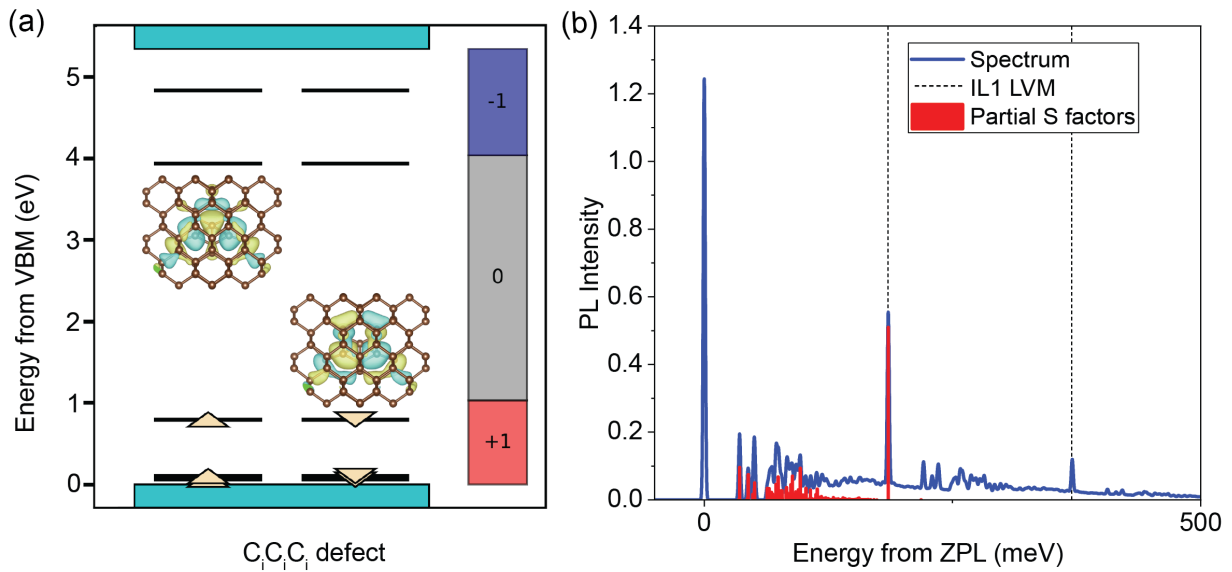

**Fig. 6.** Electronic and optical properties of the neutral $C_iC_iC_i$ defect from first-principles calculations. (a) Spin-polarized ground state defect levels inside the diamond bandgap. Occupied levels are indicated with triangles. The Kohn-Sham wavefunction is plotted for the highest occupied and lowest unoccupied orbitals using the same isovalue. Charge transition levels are indicated on the right side. (b) Simulated photoluminescence (PL) sideband with the partial contributions of the phonon modes to the Huang-Rhys (S) factor. The position of the experimental IL1 LVM is indicated by a dotted line.

### **Discussion:**

Unwanted phonon interactions constitute one of the major challenges for the development of quantum devices based on solid-state defects. At liquid helium temperatures, the equilibrium phonon populations are reduced by many orders of magnitude, but the intrinsic electron-phonon coupling processes, especially those that result in the generation of phonons, generally remain. Efforts to mitigate phonon coupling include phononic bandgap crystals[42], which suppress interactions with specific phonon modes, and nanoparticle hosts where similar effects are expected[43,44]. Additionally, phonon decoupling has been observed in interlayer complexes in 2D materials such as hBN [45].

An emerging class of phonon-decoupled defects in diamond[19–21] featuring unusually narrow linewidths and strongly suppressed phonon sidebands, suggests that previously unexplored phonon isolation mechanisms are at play (Fig. 7). The exceptional properties of IL1 centers in combination with the simulation results bring insight into the key aspects of this mechanism. The ultra-narrow ZPL linewidth in IL1s can be explained by a combination of single orbital states and small nanodiamond size. The temperature dependence of the ZPL linewidth points to two-phonon elastic scattering as the dominant broadening process whose rate scales as $T^3$. For nearly-degenerate orbital levels as in G4V centers[46], orbital mixing processes, including inelastic scattering, also contribute to the linewidth. This is in contrast with single orbital transitions, where pure dephasing dominates, leading to a smaller homogeneous ZPL broadening, such as in the case of the IL1 center. In addition, in small NDs, one may expect the suppression of phonon density of states at frequencies below $f_{min} \sim v_s/d_{ND}$, where $v_s$ is the diamond sound velocity and $d_{ND}$ is the nanodiamond radius. In our NDs with sizes of the order of 50 nm, this limit can be as high as a few hundred GHz, further suppressing the elastic two-phonon scattering.

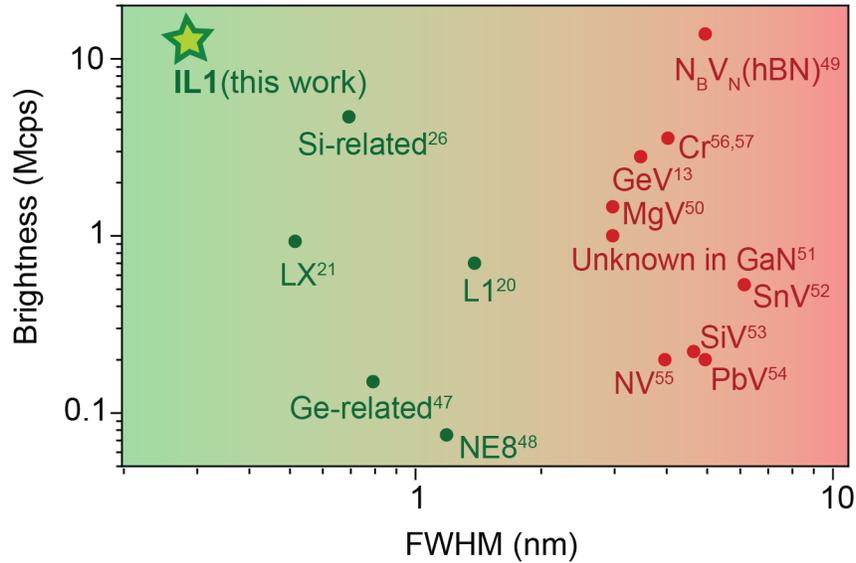

**Fig. 7.** Comparison of room-temperature optical properties of IL1 (starred) with other single narrowband SPEs[13,20,21,26,47–57]. The red dots denote SPEs with substantial bulk PSB. The green dots represent an emerging class of strongly phonon-decoupled emitters with ultranarrow linewidths and strongly suppressed bulk PSB.

To qualitatively explain bulk PSB suppression in IL1s, we turn to the Frank-Condon model of vibronic transitions. An exceptionally strong radiative decay confirmed by a substantial overlap between the excited (*e*) and ground (*g*) state electronic orbitals implies that the charge density remains largely unchanged during the optical transition. Consequently, a small shift in equilibrium atomic positions between *g* and *e* electronic orbital states limits the total expected vibrational energy transferred to the lattice during the transition. Furthermore, the transferred energy is partitioned among the phonon modes according to their degree of spatial overlap with the color center atoms. The LVM, by not coupling to the bulk modes, necessarily features the highest overlap as the atomic amplitude in the LVM peaks around the location of the color center. Thus, the LVM claims the majority of the released vibrational energy, further suppressing the bulk PSB.

The IL1 center offers exciting possibilities for device engineering and the exploration of new quantum resources. ND hosts are suitable for hybrid integration into photonic and plasmonic cavities on any material platform[9], with the benefit of speeding up the emission and improving the quantum yield. A lifetime shortening of a few hundred times may already lead to the emission of indistinguishable photons at non-cryogenic temperatures (>150K), achievable by integration into hybrid or even dielectric cavities[9,58,59].

The exclusive coupling of the electronic transition to a single high-energy localized vibrational mode in the IL1 provides a system for engineering these phonon modes to a regime relevant for applications in quantum information processing, such as quantum memories[60–62]. The non-propagating nature of localized phonons will facilitate strong, spatially confined coupling to control pulses, enabling efficient control over phonon-mediated interactions. Finally, the observed anharmonicity of the LVM states should enable direct control of single-vibron excitations.

Additionally, the strong ZPL transition in the IL1 centers can be utilized for all-optical temperature measurements relevant for biological applications. A linear fit of the experimental data in Fig. 3(c) to a line yields a slope of at least $\Delta\lambda/\Delta T=0.007$ nm/ K near room temperature range (280-300 K). Similar susceptibility is also estimated for peaks 1PL and 2PL. Furthermore, the exceptionally narrow linewidth of ZPL and the collected high brightness will allow high-sensitivity measurements at biologically compatible excitation intensities, with a noise floor on the order of $\sim 100 \text{ mK} \cdot \text{Hz}^{-1/2}$ per color center[63–65] (refer Supplementary Information, S8). Scaling this approach to IL1 ensembles can unlock all-optical thermometry with nanoscale resolution and sensitivities comparable to those achieved with ODMR-based techniques, without the technical complexities associated with microwave power delivery[66].

## Methods:

### Sample fabrication

The nanodiamonds studied in this work were synthesized by HPHT treatment of catalyst metal-free halogenated heterorganic growth system based on homogeneous mixtures of 1-fluoroadamantane, $C_{10}H_{15}F$ (BLD Pharm) and triphenylchlorosilane, $C_{18}H_{15}ClSi$ (Sigma-Aldrich), in weight ratio of 150:11. Cold-pressed tablets of the homogeneous initial mixtures (5 mm diameter and 4 mm height) were placed into a graphite container, which simultaneously served as a heater of the high-pressure apparatus. The synthesis was performed in a high-pressure apparatus of "Toroid" type at 8.0 GPa, 1400° C, and short isothermal exposure under constant load for 2 s. The obtained diamond products were then isolated by quenching to room temperature under pressure. Typical Raman spectrum obtained from the resulting powder shows a distinct diamond Raman line at 1332 cm$^{-1}$ (Fig. 1(a)). A typical TEM image of the resulting nanodiamond particles is presented in Fig. 1(b). Post-synthesis, the product is cleaned with $HNO_3$:$HClO_4$:$H_2SO_4$ at 200 °C for 48 hours to eliminate graphitic components.

Samples for optical studies are prepared by dropcasting a sonicated solution of NDs in a 1:1 water and ethanol mixture on a flat Si substrate for the temperature dependence study, and a coverslip glass for the other measurements.

### Characterization

For room temperature measurement of photophysical properties (second-order autocorrelation, saturation, lifetime, and polarization), optical characterization was performed using a custom-made scanning confocal microscope with a 100 μm pinhole based on a commercial inverted microscope body (Nikon Ti–U). A CW green laser emitting at 518 nm (Coherent OBIS 520LX) was used for optical pumping in all the experiments except the lifetime measurement. Lifetime characterization was performed using a frequency-doubled 518 nm pump from a pulsed laser (Mai Tai HP) with a nominal 100 fs pulse width and 80 MHz repetition rate. The excitation beam was reflected off a 550 nm long-pass dichroic mirror (DMLP550L, Thorlabs), and a 550 nm long-pass filter (FEL0550, Thorlabs) was used to filter out the remaining pump power. Two avalanche photodetectors with a 30 ps time resolution and 48% quantum efficiency at 550 nm (PDM, Micro-Photon Devices) were used for single-photon detection during scanning, lifetime, and autocorrelation measurements. An avalanche photodetector with a 55% quantum efficiency at 550 nm (SPCM-AQRH, Excelitas) was used for saturation measurements. Photon coincidences were measured by a time-correlated single photon counting system (Hydra Harp 400, Picoquant).

Temperature-dependent PL measurements were performed using a home-built confocal PL microscope integrated into a closed-cycle Montana S100 cryostation. A Princeton Instruments Isoplane SCT-320 spectrograph with a Pixis 400BR Excelon camera was used to measure PL spectra. A grating with 2400 grooves/mm was used to measure PL peaks in Fig. 3. with a spectral resolution of ~0.05nm.

The PL characterization for statistical studies at room temperature, including the Raman scattering, were performed on emitters exhibiting distinct harmonic spectra with 4 peaks. A WITec Alpha 300 RA Raman-AFM-SNOM system was used to measure spectra. Calibration of the spectrometer was done using an Ar/Hg spectral lamp, giving an accuracy of 0.13 cm$^{-1}$. A grating with 1800 grooves/mm was employed, providing a spectral resolution of ~ 0.06 nm. A non-resonant excitation source of 532 nm laser was focused through a microscope objective with NA = 0.9 to measure diamond Raman peaks in nanodiamond clusters on coverslip glass.

**Calculations:**

Density functional theory calculations were carried out using the VASP plane wave-based code[67–69] with a cutoff energy of 500 eV and sampling only the gamma-point in the reciprocal space. Phonon calculations for the search of high-energy LVMs were carried out in a 64-atom supercell using the PBE functional[70]. The optical properties of the candidate $C_iC_iC_i$ defect were calculated in a 512-atom supercell using the HSE06 hybrid functional [71]. The excited state was calculated with the ΔSCF method and ionic relaxation. Transition dipole moments were calculated using the PyVaspwfc code[72]. Periodic charge correction of the total energies was obtained using the FNV method [73]. The dissociation of the $C_iC_iC_i$ defect was modeled in climbing image nudged elastic band calculations at the PBE functional level using the 512-atom supercell.

**Acknowledgments**

The authors acknowledge C. Becher and A. J. Cyphersmith for useful discussions. This material is based upon work supported by the National Science Foundation NSF EAGER Grant 22-40621 and NSF CAREER Grant 22-40621. Any opinions, findings, and conclusions or recommendations expressed in this material are those of the author(s) and do not necessarily reflect the views of the National Science Foundation. The temperature dependence of spectral lines was conducted as part of a user project at the Center for Nanophase Materials Sciences (CNMS), which is a US Department of Energy, Office of Science User Facility at Oak Ridge National Laboratory. The room temperature spectral measurements were carried out in part using the Core Facilities at the Carl R. Woese Institute for Genomic Biology at UIUC. This work was supported by the U.S. Department of Energy, Office of Science, Basic Energy Sciences, Materials Sciences and Engineering Division (DOE BES MSE) and was performed using resources of the National Energy Research Scientific Computing Center, a DOE Office of Science User Facility supported by the Office of Science of the U.S. Department of Energy under Contract No. DE-AC02-05CH11231 and Award No. HEP-ERCAP0033510.

**Author contributions**

S.I.B. and S.S. conceived the study. V.A. and V.D. fabricated the NDs and carried out microRaman spectroscopy and TEM imaging. S.S. conducted the EDS, SEM, AFM, and room temperature PL measurements. J.L. performed the fluorescence lifetime, polarization and autocorrelation analysis and fitting with input from S.S. and S.I.B. D.K. measured the room temperature PL spectroscopy.

S.S. and B.L. carried out temperature-dependent PL measurements. P.U. developed the first-principles model with input from P.N. S.S., P.U. and S.I.B. wrote the manuscript. All authors helped revise the manuscript to its final form.

**Competing interests**

Authors declare that they have no competing interests.

**Data and materials availability**

The data that support the findings of this study are available from the corresponding author upon reasonable request.